\documentclass[12pt]{article}

\input epsf
\def\beq{\begin{eqnarray}}
\def\eeq{\end{eqnarray}}

\def\lsim{\mathrel{\rlap{\lower3pt\hbox{\hskip0pt$\sim$}}
     \raise1pt\hbox{$<$}}}         
\def\gsim{\mathrel{\rlap{\lower4pt\hbox{\hskip1pt$\sim$}}
     \raise1pt\hbox{$>$}}}         

\begin{document}
\begin{flushright}
SLAC-PUB-10196\\
SU-ITP-03/27\\
NYU-TH-30/10/3 

\end{flushright}

\vskip 1cm
\begin{center}
{\Large \bf
Large scale power and running spectral index in New Old Inflation
\vskip 0.2cm}
\vspace{0.4in}
{Gia Dvali$^a$ and Shamit Kachru$^b$}
\vspace{0.3in}

$^a$ {\baselineskip=14pt \it
Center for Cosmology and Particle Physics \\[1mm]
}{\baselineskip=14pt \it
Department of Physics, New York University,
New York, NY 10003\\[1mm]}

\vspace{0.2in}

$^b$ {\baselineskip=14pt \it
Department of Physics and SLAC \\[1mm]}
{\baselineskip=14pt \it Stanford University,
Stanford, CA 94305\\[1mm]}

\end{center}
\vspace{0.2cm}
\begin{center}
{\bf Abstract}
\end{center}
\vspace{0.2cm}
We have proposed a new class of inflationary scenarios in which the
first stage of expansion 
is driven by ``old" false vacuum inflation.  This ends by
nucleation of a bubble, which then further inflates.
Unlike the standard slow-roll scenarios 
the ``clock" ending the second inflationary phase 
is not a local order parameter, but rather the 
average value of an oscillating scalar
field, which locks the system at a saddle point of the potential 
in a temporary inflationary state. Inflation ends when the
amplitude drops below a certain critical point and 
liberates the system from the
false vacuum state.  The second stage of inflation has 
only about 50 e-foldings, a number which
is determined entirely by the ratio
of the fundamental mass scales, such as the
Planck/string scale and the supersymmetry breaking scale.
The density perturbations are generated due to fluctuations of moduli-dependent
Yukawa couplings. In this note we explore the observable imprints in the
fluctuation spectrum of generic cross-couplings in the superpotential and 
in the K\"ahler potential. We show that in the presence of generic 
non-renormalizable interactions
in the superpotential between the fluctuating modulus 
and the oscillating inflaton, the amplitude of the 
density perturbations is exponentially cut-off for 
sufficiently large wavelengths.  With reasonable choices of scales and
interactions, this long wavelength cutoff can occur at approximately
the current horizon size. The perturbative corrections in the K\"ahler 
potential give non-trivial potentially observable tilt and a running of the 
spectral index which is different from the standard inflationary models.

\newpage

\section{Introduction}

In a recent paper \cite{NewOld}, we suggested an inflationary scenario which builds
on Guth's old inflation \cite{Guth}.
The basic idea is to start in a false vacuum as in old inflation, and construct
our observed universe from a single bubble of false vacuum decay. 
This requires
a brief period of inflation after bubble nucleation, and suitable 
mechanisms for
reheating and production of density perturbations within the bubble. 
This cosmological sequence is similar in spirit to ``New'' 
inflation \cite{new}, but the way it is achieved in our scenario 
is dramatically different.  The key difference is that the ``clock'' that
controls the expansion of the bubble is not a local order parameter, i.e.,
the slowly rolling inflaton field.  Instead, it is 
an averaged amplitude of 
a quickly oscillating scalar field, that locks the system in an inflationary
state.  Simple toy
models which accomplish this aim
were presented in \cite{NewOld}.  They involve a rolling scalar field
$\Phi$ which stabilizes a second field $\phi$ in its false 
vacuum.\footnote{The potentials studed in \cite{NewOld} are essentially
identical to those which occur in hybrid inflation \cite{hybrid}, though
we only use the potential in a 
small-field region where slow-roll inflation would
not occur. 
With different choices of parameters than we make, these
potentials have also been studied in
connection with parametric resonance in \cite{pares}.}
The 
$\Phi$ field oscillations are redshifted away during inflation, and eventually 
$\phi$ is destabilized and rolls to its true vacuum.

 The idea of separating the inflationary ``clock'' from the velocity
of the slowly rolling scalar field 
goes back to Linde's ``Hybrid Inflation'' \cite{hybrid}, 
in which (in some regimes) the inflation can be terminated almost
instantly by triggering a phase transition in the second field. 
However, the clock that triggers this transition is still a local 
slowly-rolling field.
In this respect our attitude is more radical, since in our case what 
changes slowly is not a local scalar field, but rather its averaged
amplitude. 

The chief merit of these models is that the required 
inflationary potentials are natural from the viewpoint of softly broken
supersymmetry, and do not have to satisfy any restrictive
slow-roll conditions.
However some restrictive assumptions are required to design full models 
with appropriate reheating and density perturbations.

In particular, for the choices of scales which are natural in the models
of \cite{NewOld}, it is necessary to find a nonstandard way of
generating the observed density perturbations -- given the low scale of
inflation and the absence of slow-roll in the minimal scenario, 
the perturbations generated by
the inflaton are negligible.  The decay-rate fluctuation
mechanism of \cite{dgz,k} is ideal for this purpose, and was suggested as
an appropriate framework.  In this note, we point out that in models of
this sort, generic non-renormalizable couplings of the fluctuating
modulus field $\chi$ of \cite{dgz,k} to the rolling scalar field $\Phi$ which produces
the bubble inflation can have striking observational consequences. They 
give the $\chi$ field an effective mass $m_{\chi} > H_*$ during the early
e-foldings of the bubble inflation, leading to an absence of perturbations
on the largest scales!
The perturbations turn on as the oscillation amplitude of $\Phi$ decays 
away, and $m_{\chi}$ relaxes to $m_{\chi} << H_*$. 
This can have two interesting consequences:

\noindent
1) For reasonable choices of the $\Phi-\chi$ coupling, one can design models
where the perturbations become negligible at scales close to our present
horizon size.
CMB measurements at the largest scale are cosmic-variance limited, and
so one would like to find other consequences/predictions of such a scenario.  

\noindent
2) In the same class of scenarios, one finds that the spectral index $n(k)$
has striking behavior at the wavelengths where the $\chi$ field mass is
passing through $H_*$.  The form of $n(k)$ that one finds is unlike
that which occurs in other inflationary models, and is in principle a
testable prediction.   

The organization of this note is as follows.
In \S2, we describe the scenario of \cite{NewOld}.  In \S3, we show 
how the imprinting of density perturbations can be influenced by generic 
non-renormalizable couplings between the inflaton $\Phi$ and the fluctuating
modulus field $\chi$.  In three brief appendices, we 
address several questions
about the approximations that yield our inflationary scenario. 
We explain how the small numbers in our potentials come naturally
from soft susy breaking scenarios, we show that tunneling to nonzero
values of $\phi$ or $\dot\phi$ is highly unlikely (such paths are quickly
drawn back to the one we considered in \cite{NewOld}), and we show that
perturbative annihilation of $\Phi$ quanta has a negligible influence
on our scenario.

Several other ideas about producing 
features in the CMB at low $l$ have been proposed,
see e.g. \cite{fang,tegmark,ellis,Esf,andrei,cutoff,cline} 
and references therein.  
In particular in section IIIB of \cite{cutoff}, similar potentials
to ours are discussed in a different regime, in exploring a different idea 
about the same problem.

\section{Scenario}

The basic building block of our ``locked" inflationary scenario is a 
system of two scalar fields $\Phi$ and $\phi$ with 
an unsuppressed cross-coupling in the potential,
\begin{equation}
\label{coupl}
\Phi^2\phi^2,
\end{equation}
such that  when either of the
fields is large, the other one acquires a big mass and gets ``nailed" 
to a zero value. This is a characteristic feature of the ``hybrid'' 
model \cite{hybrid},
where this coupling traps one field in a false vacuum, while the other is
slowly-rolling. The regime that we are exploring, however,
does not rely on the slow-roll.

We assume that the self-interaction potentials $V(\Phi)$ and $V(\phi)$ are generated as a result of supersymmetry-breaking effects in the K\"ahler potential, and are such that the system has:
{\it 1)} a false minimum at some $\phi \, = \, 0, \Phi\, = \, \Phi_{false} \, \sim \,M_P$,
{\it 2)} a true minimum at $\phi\, = \, \phi_{true},\,  \Phi\, = \, 0$, and {\it 3)} a saddle point at $\phi \, = \, \Phi\,  =0$.  Inflation starts in the false vacuum, where $\phi$ has a mass $\sim M_P$, and
cannot affect the tunneling dynamics.   So $\Phi$ tunnels 
towards the saddle point, and materializes
at some initial value $\sim M_P$ or so.  Throughout this process,  
the $\phi$ field is superheavy
and is firmly fixed at zero value (since this has caused some confusion, 
see the appendix for elaboration on this point).   
After bubble nucleation, $\Phi$ rolls towards the saddle point and oscillates about it. 
These oscillations induce an effective positive mass$^2$ term
for the $\phi$ field, and prevent an instability in the $\phi$-direction from
developing. Hence the system is
locked in a temporary false vacuum state, with energy density 
consisting of the constant false 
vacuum potential energy at the saddle point, plus 
the energy density of the oscillator, which redshifts as
matter.  

Eventually, the oscillator energy density becomes sub-dominant and the system
inflates.
Since $\Phi$ is a very weakly self-coupled
field,  we shall ignore the non-linear part of its self-interactions
after tunneling and will only keep the harmonic term in the potential.
The potential then has the form
\begin{equation}
\label{pot1}
 m_{\Phi}^2 \, \Phi^2 \,  + \, \Phi^2\phi^2  \, + \, V(\phi)
\end{equation}
where $V(\phi)$ is a self-coupling potential of $\phi$, which 
has a maximum at $\phi =0$ and
a minimum at $\phi = \phi_{true}$.

 Before discussing $V(\phi)$, let us briefly discuss corrections 
to the potential of $\Phi$. 
Since inflation breaks supersymmetry (spontaneously), the perturbative 
corrections to the K\"ahler potential arising from $\phi$-loops will 
correct the potential of $\Phi$, even if all other corrections are 
absent \cite{dss}. The resulting one-loop Coleman-Weinberg potential for 
$\Phi \, >> \, m_{soft}^2$ behaves as  
\begin{equation}
V_{one-loop} \, \simeq \, {m_{soft}^2 \over 
32\pi^2} |\Phi|^2 {\rm ln}(|\Phi|),
\end{equation}
where  $m^2$ is a
soft mass of $\phi$. The cancellation of the $\Phi^4$ term among bosons 
and fermions is a general consequence of supersymmetry, which 
holds even though we evaluate corrections along the 
inflationary trajectory where SUSY is spontaneously broken \cite{dss}.   
It is clear that in our case, these correction are so small that
they are unimportant 
for the inflationary dynamics.  However, as we shall see, 
similar corrections 
may play an important role in creating a non-trivial tilt in the spectrum 
of perturbations. 

We shall assume that $V(\phi)$ is a typical potential for a field which
is a supersymmetric flat direction, whose potential
comes entirely from the K\"ahler potential after supersymmetry-breaking.  The corrections to
the K\"ahler potential that induce the $\phi$ VEV may come either from tree-level 
gravity-mediated susy-breaking
or from perturbative  renormalization due to matter loops. In 
the former case the VEV at the true minimum is typically $\phi_{true} \sim M_P$, whereas 
in the latter case we may have
$\phi_{true} << M_P$.  Both possibilities are considered in the appendix. 
Irrespective of the precise
form of these corrections, for the flat direction fields whose potentials
and preferred VEVs are generated as a result of
soft supersymmetry breaking, the following is true in general.  The 
curvature at $\phi = 0$
is $\sim  m_{soft}^2 $, and the value of the false vacuum energy is
\begin{equation}
\label{false}
V(0) \, = \, m^2_{soft}({{\phi_{true}^{p}}\over{{M_P}^{p-2}}}) 
\end{equation}
for some $p$.  For instance for the model discussed in \S2\ of \cite{NewOld},
one has $p=4$.  

The inflationary Hubble parameter in such a model is given by
$H_* \sim m_{soft} \, ({\phi_{true} \over M_P})^{p/2}$.
The bubble size today is:
\begin{equation}
\label{rtoday}
R_{today} \, \sim \, \, c^{{5p\over 6}} \, \left ( {M_{P} \over H_*} \right )^{{7\over 6}}\, {1 \over T_{today}}
\end{equation}
where $c \, = \, \phi_{true}/M_P$.  This expression was derived as follows.  The initial value
of the $\Phi$ field inside the bubble is $\Phi_{in}\sim M_P$.  Hence the initial energy density
is $ m_{soft}^2M_P^2$, and the initial Hubble $H \sim m_{soft}$, which 
also sets the size of
most generic bubble to be $\sim \, 1/m_{soft}$. If
$\phi_{true} << M_P$, initially the energy of $\Phi$-oscillations dominates and there is an interval of
matter-dominated expansion until the energy of oscillations becomes sub-dominant to
$V(0)$ (that is until the amplitude of $\Phi$ becomes less than $M_P c^{p/2}$).
 During the matter-dominated interval the bubble interior grows by a factor 
$c^{-{p\over 3}}$, 
after which inflation begins and the universe begins to exponentially expand.
The expansion factor during the subsequent phase of locked inflation is given by
\begin{equation}
\label{N}
e^{N} \, = \, \left ( c^{p/2 - 1}{\phi_{true} \over m_{soft}} \right )^{{2 \over 3}}.
\end{equation}
Finally there is an additional expansion factor after reheating
\begin{equation}
\label{rehexp}
{T_R \over T_{today}} \, \sim \, {\sqrt{H_*M_P} \over T_{today}}
\end{equation}
Combining all the factors, we find equation (\ref{rtoday}).

\section{Density perturbations}

The density perturbations are 
generated through the ``decay-rate-fluctuation" mechanism of
references \cite{dgz,k}. The idea is that the decay rate $\Gamma$ of the 
field $\phi$ 
(or some 
other field responsible for the reheating, as in \S4.2 of \cite{NewOld}) is
controlled by a fluctuating modulus field.  This is reasonably motivated 
by string theory, 
where couplings of the low energy fields are set by the expectation values of moduli.
Let $\chi$ be a modulus that controls the decay rate of $\phi$.  
If the mass of $\chi$ is an order of magnitude smaller
than $H_*$, fluctuations will be imprinted in $\chi$ during inflation.  These
$\chi$ fluctuations will
translate into density perturbations because they will lead to
fluctuations in the $\phi$ decay rate during reheating. The resulting
density perturbations are given by
\begin{equation}
\label{delta}
{\delta \rho \over \rho} \, =  \, - \, {2\over 3} \, {\delta \Gamma \over \Gamma}
\end{equation}
Because $\delta \Gamma \, \propto \, \delta \chi$, the 
spectrum of perturbations will be set by
the spectrum of $\chi$ fluctuations.
  
Our main point is the following.
In the presence of generic Planck scale suppressed interactions between $\chi$
and the oscillating inflaton field, the resulting spectrum 
is very peculiar and exhibits a sharp cut-off
at large wavelengths, which is potentially observable.
Imagine that $\chi$ and $\Phi$ have the 
following generic coupling in the potential, arising from a coupling
in the superpotential: 
\begin{equation}
\label{nonr}
{\Phi^n \over M_P^{n-2}}  \, \chi^2
\end{equation}
Then, the fluctuations in $\chi$ at the beginning of inflation will be strongly suppressed, due to its
high effective mass
\begin{equation}
\label{mueff}
\mu_{eff}^2 \, = \, {\langle \Phi_{in}^{n} \rangle  \over M_P^{n-2}}\, e^{-{3\over 2}nN} \, + \, \mu^2,
\end{equation}
where $\langle \Phi \rangle$ is the amplitude of oscillations and $N$ is
the number of e-foldings since the start of inflation.
$\mu$ is the ``bare'' mass of $\chi$, which for us means the full 
contribution to the mass arising from the 
the K\"ahler potential (generally field-dependent).  We 
assume it to be somewhat below $H_*$.
Fluctuations of $\chi$ then are governed by the following equation
\begin{equation}
\label{chifl}
\ddot{\delta\chi_k} \, + \, 3\, H\, \dot{\delta\chi_k} \, + \, 
(\, \mu_{eff}^2 \, + \, k^2/a^2\,)\, \delta\chi_k \, = \, 0  
\end{equation}
 So $\chi$ can only start
to fluctuate after its effective mass
drops below $\mu_c \simeq H_*$, which happens
only some critical number $N_c$ of e-foldings after the onset 
of inflation. 
Hence,
density perturbations should be cut-off at large scales.

To estimate $N_c$ recall that inflation begins when the amplitude of 
$\Phi \,\sim \, M_P c^{p/2}$,
and after this point it decays as $\langle \Phi \rangle \, = \, M_P c^{p/2} \, e^{-{3\over 2}N}$. Thus, we find
\begin{equation}
\label{ncrit}
e^{-N_c} \, \sim \, \left ({H_* \over M_P}\right ) ^{{4 \over 3n}} \, c^{-{p\over 3}}
\end{equation}
Correspondingly the maximal wavelength beyond which perturbations will be suppressed is given by
\begin{equation}
\label{ }
k_c^{-1} \, \sim \,   \, c^{{p\over 3}} \, \left ( {M_{P} \over H_*} \right )^{{7\over 6}\, - \, {4 \over 3n}} {1 \over T_{today} }
\end{equation}

 Taking $c=1$ and $p=4$ (as in \cite{NewOld}), 
with $H_* \sim 10^{-12}$ GeV (to allow TeV reheating), 
this is roughly the size of the present
horizon ($\sim 10^{28}$ cm) for $n=6$. 
For the more generically expected $n=4$, one would have to choose smaller
values for $H_*$ and/or $c$ 
to accommodate a sufficiently
small wavelength to be relevant to observations in our horizon.

It is well known that cosmic variance bounds make it difficult to improve our
certainty about the significance of the observed low quadrupole and octupole in the
CMB.  However, the measurements of the spectral index $n(k)$ will improve in
the future.  With this in mind,  
let us now discuss the $k$-dependence of the cut-off. The 
perturbations generated at $N < N_c$
(that today have $k < k_c$) are suppressed as
\begin{equation}
\label{deltan}
\delta_{N \, < \, N_c} \sim H_* \, e^{{- 2\pi \mu(t) \over H_*}}\, = \,  
H_* \, e^{- 2\pi \,e^{ -(N-N_c) { 3n \over 4}}}
\end{equation}
This can be simply understood: de Sitter space has a Gibbons-Hawking
temperature ${H_* \over 2\pi}$ \cite{GH}, and (\ref{deltan}) is 
just a reflection
of the Boltzmann-suppressed excitations of a massive field at this
temperature.
Hence today we should see a 
sharp cut-off in the perturbation amplitude at large scales with the
following $k$-dependence
\begin{equation}
\label{cutof}
\sim e^{-2\pi \left ({k_c\over k}\right )^{{3n\over 4}}}.
\end{equation}
This is a very sharp cut-off even for the lowest possible 
choice of $n=4$, and implies an abrupt
decline in the spectrum for $k\, < \,k_c$.

The fit of the observed data to a perturbation spectrum with a sharp cut-off 
was done in reference \cite{cutoff},  
according to
which the best 
fit occurs for $k_c^{-1} \sim 10^{28}$cm.  Although our cutoff is 
smoother than the one postulated
by the authors of \cite{cutoff}, from the existing analysis 
it seems difficult to distinguish between the
two possibilities, and more precise studies are needed.

Because our cut-off is so sharp for $k<k_c$, it would be difficult
to distinguish this mechanism from others which produce a sharp cutoff
based on that feature alone.
However, we also have a very different
prediction for the behavior of the spectral index $n(k)$ on scales $k>k_c$.
In particular, the way the tilt is imprinted in the spectral index
is very different from the standard inflationary case.
In slow roll inflation, the primary source for the tilt is the
fact that the Hubble parameter inevitably changes during the last
$60$ or so e-foldings, resulting in a subsequent change of the
perturbation amplitude (this is the reason that in 
standard inflationary models, having $n=1$ is unnatural).
For a thorough discussion of the perturbations in standard
inflation, see e.g. \cite{mukh}.
In our scenario, the Hubble parameter is essentially constant throughout
the inflationary period and its evolution cannot result in a tilt.  A tilt
is nevertheless generated during the evolution of $\chi$-fluctuations on 
superhorizon scales, due to the simple fact that different wavelengths
spent different amounts of time before finally being translated into 
density perturbations by reheating. 

After a given wavelength of 
$\chi$ crosses outside the inflationary horizon (or more precisely, after
 $k/a\, < \, 1/\mu$), the $k^2$-term in equation (\ref{chifl}) becomes 
subdominant
and the mode  is almost frozen, apart from the fact that the small
``bare'' mass term $\mu^2$ very slowly pushes it down.
As stated above, this mass term
comes from the K\"ahler potential, and is 
assumed to be somewhat below $H_*$. Because $\chi$ and $\Phi$ are most likely
coupled through the K\"ahler potential, $\mu$ is 
not in general a constant throughout  
the inflation, but rather undergoes some change itself.

For $\mu^2(t)\, << \, H_*^2$, the $\ddot{\delta\chi_k}$ term in 
equation (\ref{chifl}) is subdominant, and the evolution 
of the fluctuations on 
superhorizon scales is given by the following equation: 
\begin{equation}
\delta\chi_k \, \propto \, {\rm e}^{-\,\int_0^t\, {\mu^2(\tau) \over 3 
H_*}\, d\tau}
\label{sol1}
\end{equation}
Remembering that $k \, \propto \, {\rm e}^{-H_*t}$, this can be recast as
\begin{equation}
\delta\chi_k \, \propto \,  k^{{1 \over t}\,\int_0^t\, {\mu^2(\tau) \over 
3 H_*^2}\,d\tau} 
\label{sol2}
\end{equation}
This implies that the spectral index is given by 
\begin{equation}
n\, - \, 1 = \, {1 \over t}\,\int_0^t\, {\mu^2(\tau) \over 3 H_*^2}\, d\tau 
\label{n-1}
\end{equation}
and for the running we have
\begin{equation}
{dn \over d\, {\rm ln}k} \, = \, {1 \over ({\rm ln}k)^2}\,\int_0^t\, 
{\mu^2(\tau) \over 3 H_*}\,d\tau\,
+ \, {1 \over {\rm ln}k}\, {\mu^2(t) \over 3 H_*^2} 
\label{dn}
\end{equation}
For the particular case of a constant $\mu^2$ we would immediately get
\begin{equation}
n\, - \, 1 \, = \, {\mu^2 \over 3 H_*^2}
\label{nc}
\end{equation}

However, in general $\mu(t)$ won't be constant, even after the contribution 
from the cross couplings in the superpotential become negligible. This 
is because of the cross couplings between $\chi$ and $\Phi$ in 
the K\"ahler potential.  It follows from the standard rules for 
supersymmetric Lagrangians that these couplings are 
universally suppressed by the soft supersymmetry breaking scale
(the K\"ahler potential couplings can only 
contribute in the scalar potential if the 
auxiliary 
$F$-component of at least one of the chiral superfields is non-zero, see 
the appendix for a detailed discussion).  In our case the susy breaking 
scale  is $\sim H_*^2$, and so the corrections to  $\mu^2$ coming from the 
K\"ahler potential can generically be parameterized as 
\begin{equation}
\mu^2 \, = \, H_*^2 \, \left ( \alpha_0 \, + \, \alpha_1\, {\rm 
ln}(|\Phi|) \, + \, \alpha_2
\, ({\rm ln}(|\Phi|))^2\, + \, \cdots + \,  (|\Phi|/M_P)^n  + \cdots \right ) 
\label{muexp}
\end{equation}
where we have explicitly separated the $\Phi$-independent ($\alpha_0$-term) 
and the different types of 
$\Phi$-dependent contributions.  The $\alpha$-coefficients parameterize their 
strength relative to the tree-level soft supersymmetry breaking 
measured by $H_*$. 

Because the amplitude of $\Phi$ decreases exponentially quickly, the 
power-law terms very quickly diminish after the onset of inflation, and 
cannot produce any observable effect. The log-terms in contrast change 
very slowly, and have a potentially observable late time effect. So we 
shall now focus on these terms. Their origin is the {\it perturbative} 
renormalization of the K\"ahler potential. Because, unlike the 
superpotential,  
the K\"ahler metric is not protected by any non-renormalization theorem, 
it 
is unnatural to assume any strong suppression of the $\alpha$-coefficients. We 
shall take their typical value to be roughly a one or two-loop factor times
a number of order one.  
Taking into account the explicit time-dependence of the $\Phi$-amplitude, 
and using equations (\ref{n-1}) and (\ref{dn}), we arrive at the following 
expressions for the tilt 
\begin{equation}
n\, - \,1  \, = \, {\alpha_ 0\over 3 } \, + \, {\alpha_1\over 4}{\rm ln}{k 
\over k_0} \, + \, {\alpha_2\over 4} 
{\rm ln}^2{k\over k_0} 
\label{n-11}
\end{equation}
and the running
\begin{equation}
{dn \over d\, {\rm ln}k} \, = \, {\alpha_ 1\over 4} \, + \, {\alpha_2 
\over 2} {\rm ln}{k \over k_0}
\label{dn1}
\end{equation}
The  $k$-dependence of the above cosmological parameters is rather 
striking and different from the known inflationary models.  The 
precise values of the $\alpha$-coefficients are model dependent, but their 
natural values are consistent with the present WMAP data and are 
potentially detectable in the future. According to 
this data \cite{wmap}, for 
instance, for  $k_0 \, = \, 0.002$ Mpc$^{-1}$, the central value of our 
$\alpha_1\over 2$ is around $0.077$.  This is certainly consistent with 
the natural assumption that $\alpha_1$ is a one-loop factor.

\begin{appendix}
\section{Appendix}
\subsection{Moduli potentials}

 Let us briefly discuss the origin of $V(\phi)$.  According to our assumption, $\phi$  is some
supersymmetric flat direction field, and thus $V(\phi)$ should vanish in an exact susy limit.
Supersymmetry breaking effects transmitted through the K\"ahler potential 
induce a non-zero $V(\phi)$ with a minimum at 
$\phi = \phi_{true}$.  We shall distinguish two possibilities. 
The first is when the relevant K\"ahler
terms come from tree-level gravity mediated supersymmetry breaking. In such a case 
the K\"ahler
metric can be taken to be
\begin{equation}
\label{kahler}
K \, = \, \Theta^+\Theta \, + \, \Theta^+\Theta \, f \left({\phi \over M_P}\right ) \, + \, ...
\end{equation}
Here $\Theta$ is the ``spurion" superfield whose $F$-term breaks supersymmetry
$\Theta_F \, = \,  M^2$, and $f$ is some generic polynomial function
\begin{equation}
\label{f}
f \, = \, \lambda_1 \, {\phi^2 \over M_P^2} \, + \, \lambda_2 \, {\phi^4 \over M_P^4} +...
\end{equation}
with $\lambda_i \, \sim 1$.   After substituting the value of this $F$-term into (\ref{kahler}),  we
generate the following effective potential for $\phi$, at the 
leading order in the expansion in powers of
$f$
\begin{equation}
\label{keff}
V(\phi) \, \simeq \, - \, M^4\, f \left ( {\phi \over M_P} \right )
\end{equation}
Note the over-all minus sign in the potential relative to the K\"ahler potential.  
If $\lambda_1\, > \,0$, the potential
has a maximum at $\phi \, = \, 0$, as we need in our inflationary scenario. 
Since the higher order
terms only become significant for $\phi \, \sim \, M_P$,
the minimum of the potential 
will be established only at some large value $\phi_{true} \sim M_P$.

If, however, $\phi$ has some unsuppressed interactions with other gauge or chiral
superfields, the minimum can be established at values much smaller than $M_P$.
The reason is that the perturbative renormalization of the K\"ahler potential can dominate
over the gravity-mediated higher  corrections. The classic example of 
spontaneous symmetry breaking due to perturbative K\"ahler renormalization
is Witten's ``inverse hierarchy'' mechanism \cite{witten}. 
Another simple example is the flip of the sign of the 
gravity-mediated soft mass for $\phi$ due to perturbative running. The 
effective potential for $\phi$
then can be approximated as
\begin{equation}
\label{runningn}
V(\phi) \, = \, m_{soft}^2 \, \phi^2 \, (1 \, + \, \alpha \, {\rm ln}(\phi/M_P)) \, + \, M_P- {\rm suppressed~corrections}
\end{equation}
where $\alpha$ is the renormalization loop factor, 
and $m_{soft}^2 \, = \, - \lambda_1\,  M^4/M_P^2$ is the leading term in the
gravity mediated contribution in (\ref{keff}). Choosing $\lambda_1 \, < \, 0$,
the minimum at $\phi \, =\, 0$ will only get destabilized 
after renormalization effects are included.
For $\alpha << 1$ the minimum then will be developed at
\begin{equation}
\label{logmin}
\phi_{true} \, \sim \, e^{-{2\over \alpha} }\, M_P
\end{equation}
This gives us a natural mechanism for designing models that have $c < 1$.

\subsection{Off-trail tunneling}

 We are assuming that for the generic bubble the initial condition of the system after tunneling is
 $\Phi\, = \, \Phi_{in}, \phi = 0$. This is well justified for the following reason.
When the 
system sits in the false vacuum state $\Phi\, = \Phi_{false} \sim M_P, ~ \phi = 0$, the
curvature of the potential in the $\phi$ direction is $\sim M_P^2$, and in 
the $\Phi$ direction it is only
$\sim H^2$.  The energetically most favorable trajectory goes through the barrier, which 
is the maximum of the self-interaction potential of $\Phi$.   In the 
generic bubble,  $\Phi$ cannot tunnel all the way to
$\Phi =0$, but 
instead will be materialized at some $\Phi_{in} \sim M_P$ on the other side of the barrier.
In the language of tunneling, this is because the relevant instanton is the Hawking-Moss
instanton \cite{Hawking} which tunnels to the top of the barrier, not the true vacuum.
In the language
of stochastic inflation \cite{Linde}, this is 
because one fluctuates to the top of the hill and
then rolls down, instead of fluctuating directly to the true vacuum.

Hence everywhere around the tunneling trajectory of interest $\phi$ has a Planck scale mass.
Since the energy scales involved in the tunneling dynamics are much smaller, it 
is obvious that
$\phi$ can be integrated out and cannot influence the tunneling dynamics.
However,  let us try to keep the track of $\phi$ explicitly, and see 
what happens if the system instead
tunnels to some point  $\Phi_{in} \sim M_P, ~  \phi_{in} \neq 0$.  
Since $\phi$ has a Planckian mass,
tunneling to such a point would be equivalent to 
exciting a condensate of $\phi$ particles with
an occupation number $N_{\phi}  \sim M_P\, \phi_{in}^2$, and an energy density
\begin{equation}
\label{phie}
\rho_{\phi} \, \sim \, M_P^2\phi_{in}^2 .
\end{equation}
Unless $\phi_{in} \, < \, H_*$, the resulting energy density is bigger than the inflationary energy density,
and nucleation of such a bubble will be suppressed by an additional exponential factor
$\sim {\phi_{in} \over H_*}$.   On the other hand if $\phi_{in} \, < \, H_*$, the system will very quickly
be pulled back to the point $\phi \, =\, 0$ due to an almost 
instant decay of the $\phi$-condensate.
Indeed, the decay rate of such a condensate into $\Phi$-particles is
$\Gamma \, \sim \, m_{\phi} \, \sim \,  M_P$ (given the Planckian expectation
value of $\Phi$).  This is much faster than one inverse oscillation time in the 
$\Phi$-direction
(and also much faster than the expansion rate of the Universe $H_*$).  
So $\phi_{in}$ will decay to $\phi=0$ well before 
the system has any chance to reach the saddle point.
Similar remarks obviously apply to tunneling to a configuration with 
$\dot\phi \neq 0$.

\subsection{Decay rate of $\Phi$}

One might worry that perturbative annihilation of $\Phi$ particles to $\phi$ particles
could lead to an earlier end of the locked inflation than the naive estimate
suggests.  Here we argue that in fact, these decays are negligible.

The naive perturbative decay rate $\Phi\, \Phi \rightarrow \phi\, \phi$ is
\begin{equation}
\label{phirate}
\Gamma_{2\Phi \rightarrow 2\phi} \, \sim \, m_{\Phi}
\end{equation}
(the dimensionless coupling is set to one).
This is only applicable, however, for small oscillation amplitudes.  It 
does not take into account
the non-perturbative effect of the $\Phi$ condensate on the $\phi$ mass, 
which blocks the process.
This can be understood as follows. During most of the oscillation time $\phi$ is much heavier than
$\Phi$, and annihilation is blocked. The process is allowed in a 
very narrow time interval (per each oscillation), during which
$\phi$ is lighter than $\Phi$. The corresponding fraction of time 
per oscillation when the annihilation is allowed 
is ${m_{\Phi} \over \langle \Phi \rangle}$.
Therefore to first approximation, we can
model the system by saying that the annihilation rate will be suppressed to roughly 
\begin{equation}
\label{phiratet}
\Gamma_{2\Phi \rightarrow 2\phi} \, \sim \, {m_{\Phi}^2 \over \langle \Phi \rangle}
\end{equation}
This is negligible compared to the expansion rate of the Universe $H_*$, and is 
highly inefficient.

\end{appendix} 

\vspace{0.3in} \centerline{\bf{Acknowledgements}}
\vspace{0.2in}
We thank J. Cline, S. Dimopoulos, G. Gabadadze,
A. Gruzinov, A. Linde, J. Maldacena, V. Mukhanov,  
R. Scoccimarro, S. Shenker, A. Toporensky, 
D. Wands, E. Witten and M. Zaldarriaga for helpful discussions. 
The research of G.D. is suported in part by a David and Lucile
Packard Foundation Fellowship for Science and Engineering, and by
NSF grant PHY-0070787. The research of S.K. is supported in part by a David
and Lucile Packard Foundation Fellowship for Science and
Engineering, NSF grant PHY-0097915, and the DOE under contract
DE-AC03-76SF00515.


\begin{thebibliography}{99}

\bibitem{NewOld}
G. Dvali and S. Kachru, ``New Old Inflation,'' hep-th/0309095.

\bibitem{Guth}
A. Guth, ``The Inflationary Universe: A Possible Solution to the
Horizon and Flatness Problems,'' Phys. Rev. {\bf D23} (1981) 347.

\bibitem{new}
A. Linde, ``A New Inflationary Universe Scenario: A Possible Solution
of the Horizon, Flatness, Homogeneity, Isotropy and Primordial
Monopole Problems,'' Phys. Lett. {\bf B108} (1982) 389;
A. Albrecht and P. Steinhardt, ``Cosmology for Grand Unified
Theories with Radiatively Induced Symmetry Breaking,'' Phys. Rev.
Lett. {\bf 48} (1982) 1220.

\bibitem{hybrid}
A. Linde, ``Hybrid Inflation," Phys. Rev. {\bf D49} (1994) 748,
astro-ph/9307002.

\bibitem{pares}
See e.g.: G.N. Felder, J. Garcia-Bellido, P.B. Greene, L. Kofman,
A.D. Linde and I. Tkachev, ``Dynamics of Symmetry Breaking and
Tachyonic Preheating,'' Phys. Rev. Lett. {\bf 87} (2001) 011601,
hep-th/0012142; G.N. Felder, L. Kofman and A.D. Linde, ``Tachyonic
Instability and Dynamics of Spontaneous Symmetry Breaking,''
Phys. Rev. {\bf D64} (2001) 123517, hep-th/0106179; and
references therein.

\bibitem{dgz}
G. Dvali, A. Gruzinov and M. Zaldarriaga, ``A New Mechanism for
Generating Density Perturbations from Inflation,'' astro-ph/0303591;
G. Dvali, A. Gruzinov and M. Zaldarriaga, ``Cosmological Perturbations
from Inhomogeneous Reheating, Freeze-Out, and Mass Domination,''
astro-ph/0305548; M. Zaldarriaga, ``Non-Gaussianities in Models
with a Varying Inflaton Decay Rate,''  astro-ph/0306006.

\bibitem{k}
L. Kofman, ``Probing String Theory with Modulated Cosmological
Fluctuations,'' astro-ph/0303614.

\bibitem{fang}
Y.P. Jing and L.Z. Fang, ``An infrared cutoff revealed by the two
years of COBE-DMR observations of cosmic temperature fluctuations,''
Phys. Rev. Lett. {\bf 73} (1994) 1882, astro-ph/9409072.

\bibitem{tegmark}
M. Tegmark, A. de Oliveira-Costa and A. Hamilton, ``A high resolution
foreground cleaned CMB map from WMAP,'' astro-ph/0302496. 

\bibitem{ellis}
J.P. Uzan, U. Kirchner and G.F. Ellis, ``WMAP data and the
curvature of space,'' astro-ph/0302597. 

\bibitem{Esf}
G. Efstathiou, ``Is the Low CMB Quadrupole a Signature of Spatial
Curvature ?,'' astro-ph/0303127.

\bibitem{andrei}
A. Linde, ``Can we have inflation with $\Omega > 1$?,'' 
astro-ph/0303245. 

\bibitem{cutoff}
C. Contaldi, M. Peloso, L. Kofman and A. Linde, ``Suppressing the lower
Multipoles in the CMB Anisotropies,'' astro-ph/0303636.

\bibitem{cline}
J. Cline, P. Crotty and J. Lesgourgues, ``Does the small CMB
Quadrupole Moment Suggest New Physics?,'' astro-ph/0304558.

\bibitem{dss}
G. Dvali, Q. Shafi and R. Schaefer, ``Large Scale Structure and
Supersymmetric Inflation without Fine Tuning,'' Phys. Rev. Lett.
{\bf 73} (1994) 1886, hep-ph/9406319.

\bibitem{mukh}
V.F. Mukhanov, H.A. Feldman and R. Brandenberger, ``Theory of
Cosmological Perturbations. Part 1. Classical Perturbations.  Part 2.
Quantum Theory of Perturbations. Part 3. Extensions,''
Phys. Rept. {\bf 215} (1992) 203.

\bibitem{Hawking}
S. Hawking and I. Moss, ``Supercooled Phase Transitions in the
Very Early Universe,'' Phys. Lett. {\bf B110} (1982) 35.

\bibitem{GH}
G. Gibbons and S. Hawking, ``Action Integrals and Partition Functions
in Quantum Gravity,'' Phys. Rev. {\bf D15} (1977) 2752.

\bibitem{wmap} H.V. Peiris et al, ``First Year Wilkinson Microwave 
Anisotropy Probe (WMAP) Observations: Implications for Inflation,'' 
astro-ph/0302225.


\bibitem{witten}

E. Witten,  ``Mass Hierarchies in Supersymmetric Theories,'' 
 Phys. Lett. {\bf B105} (1981) 267.

\bibitem{Linde}
See e.g. the
discussion in:  A. Linde, {\it Particle Physics and Inflationary Cosmology}, Harwood, 1990.



\end{thebibliography}
\end{document}